\documentclass[showpacs,twocolumn,pre]{revtex4}
\usepackage{amssymb}
\usepackage{graphicx}
\usepackage{dcolumn}
\usepackage{bm}

\def\be{\begin{equation}}
\def\ee{\end{equation}}
\def\bea{\begin{eqnarray}}
\def\eea{\end{eqnarray}}
\def\ba{\begin{array}}
\def\ea{\end{array}}

\def\b{\beta}

\def\vep{\varepsilon}
\def\k{\kappa}
\def\x{\xi}

\def\0{$\Gamma_0$}
\def\o{\omega}

\def\pa{\partial}

\def\s{\sigma}
\def\m{\mu}
\def\l{\lambda}

\begin{document}

\title{Slow light, open cavity formation, and large longitudinal electric field 
on slab waveguide made of indefinite permittivity materials}
\author{W. T. Lu}
\email{wentao09@gmail.com}
\author{S. Sridhar}
\affiliation{Department of Physics and Electronic Materials Research Institute,
Northeastern University, Boston, Massachusetts 02115, USA}
\date{\today}

\begin{abstract}
The optical properties of slab waveguides made of indefinite permittivity ($\vep$) 
materials (IEM) are considered. In this medium the transverse permittivity is negative 
while the longitudinal permittivity is positive.
At any given frequency the waveguide supports an infinite number of transverse 
magnetic (TM) eigenmodes. For a slab waveguide with a fixed thickness, 
at most only one TM mode is forward-wave. 
The rest of them are backward waves which can have very large phase index. 
At a critical thickness, the waveguide supports degenerate forward- and 
backward-wave modes with zero group velocity. Above the critical thickness, 
the waveguide supports complex-conjugate decay modes instead of propagating modes.
The presence of loss in IEMs will lift the degeneracy, resulting in modes with
finite group velocity. Feasible realization is proposed. The performance of IEM
waveguide is analyzed and possible applications are discussed which are supported 
by numerical calculations.
These slab waveguides can be used to make optical delay lines in optical buffers to 
slow down and trap light,  to form open cavities, to generate strong longitudinal 
electric fields, and as phase shifters in optical integrated circuits. 
\end{abstract}


\maketitle

\section{Introduction}

There is a strong interest in slow light structures and devices 
\cite{Hau99,Biglow03,Yanik04,Camacho07,XiaF07,Baba08,Thevenaz08,Tsakmakidis,Huang08,GanQ08,GanQ09,Lu09a,Lu09b,Khurgin}. 
This is driven by the need for optical buffers and optical memory 
in optical integrated circuits and other applications.
In order to have small group velocity, various methods have been exploited to 
engineer material and structural dispersions,
such as electromagnetic induced transparency \cite{Hau08}, 
coupled resonant structures \cite{XiaF07,Notomi08}, 
and optical nonlinearity \cite{Yanik04}.

Tsakmakidis et al \cite{Tsakmakidis} proposed a novel scheme to realize trapped rainbow
by using negative-index metamaterials. This slow light waveguide requires
its core layer to be made of
double-negative metamaterial (DNM) whose permittivity and permeability are both negative.
It is quite a challenge to realize DNMs with very low loss \cite{Shalaev} since even moderate
loss will destroy the zero group velocity mode \cite{Reza08,Tsakmakidis08}. Strictly speaking, 
when loss is present as is inevitable in passive systems, storing light indefinitely is impossible. 
However, gain may be introduced to compensate loss and makes zero group velocity possible \cite{Lu09a}. Slow light waveguide with single-negative materials
has also been proposed \cite{Lu09a,Lu09b}.

In this paper we consider planar waveguide made of 
the so-called indefinite medium \cite{Smith03,Hoffman,Lu07,YaoJ} 
whose permittivity and/or the permeability tensors are indefinite matrices.
For an indefinite medium,
the dispersion is hyperbolic for one polarization and elliptical for the other.
Negative refraction, superlens imaging, and hyperlens focusing \cite{LiuZ,Smolyaninov} 
can be realized by using indefinite permittivity ($\vep$) materials (IEMs). 
Nanowire waveguide made of IEM has been considered by Huang et al \cite{Huang08}. 
These IEM waveguides can support both forward- and backward-wave modes.
High phase index can be obtained for these guided modes. 
Here we further exploit the unique properties of IEM waveguide by 
considering a more manageable planar waveguide geometry.
We will reveal that these waveguides can also support degenerate modes
which can be used to slow down and trap light, form open cavities and 
generate strong longitudinal electric field.
These unique properties can lead to a broad range of applications.

The paper is organized as follows.
In Sec. II, we present the solutions for the TM modes supported by an anisotropic 
slab waveguide.
The realization of IEMs will be proposed and light coupling into IEM waveguides  
will be discussed in Sec. III.
In Sec. IV, various unique features such as slow light, open cavity formation,
and large longitudinal electric field
will be revealed. We conclude in Sec. V.

\section{Wave propagation in an anisotropic planar waveguide}

\subsection{Eigen modes equation}
\begin{figure}[htbp]
\center{
\includegraphics [angle=0, width=6cm]{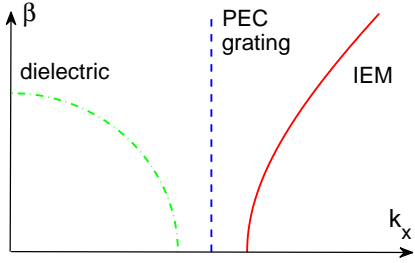}}
\caption{(Color online) Hyperbolic dispersion relation for the TM waves in an IEM.
The dashed line (blue) is for a perfect electronic conductor (PEC) grating whose transverse 
component of the wave vector is independent
of the longitudinal component $\beta$.}
\label{fig-disp}
\end{figure}

We consider a slab of planar waveguide made of anisotropic metamaterial in air.
The wave propagation is along the $z$-direction with phase $e^{i(\b z-\o t)}$ 
and the transverse direction is in the $x$-direction. 
We consider the case of an IEM with
\be
\vep_z>0,\quad \vep_x<0. \label{IEM-xz}
\ee
For the transverse magnetic (TM) modes, the magnetic field is in the $y$-direction.
The transverse component $k_x$ of the wave vector inside the metamaterial is
\be
k_x=\sqrt{\vep_z}\sqrt{\mu_yk_0^2-\b^2/\vep_x}. \label{hyper-disp}
\ee
Here $k_0$ is the wave number in the vacuum. 
Since we do not consider magnetic materials, we set $\mu_y=1$. 
Depending on the value of $\vep_y$, this extremely anisotropic medium can 
be uniaxial or biaxial \cite{Lindell,Mackay,Depine}. 
However, since $\vep_y$ is not involved with the TM modes, 
its value is not of our interest in this paper.
Due to the negativity of $\vep_x$, the dispersion Eq. (\ref{hyper-disp})
is hyperbolic instead of elliptic,
which is shown in Fig. \ref{fig-disp}.

Since the planar waveguide is symmetric, the magnetic field is
\bea
H_y(x)&=&e^{\k_0(x+d/2)},\quad x\leq -d/2,\nonumber \\
&=&Ae^{ik_xx}+Be^{-ik_xx},\quad -d/2\leq x\leq d/2,\nonumber \\
&=&e^{-\k_0(x-d/2)},\quad x\geq d/2.
\eea
Here $d$ is the slab thickness and $\k_0=\sqrt{\b^2-k_0^2}$ is the decay constant 
in the transverse direction in air.
The tangential electric field is
\bea
E_z(x)&=&{i\k_0\over k_0}e^{\k_0(x+d/2)},\quad x\leq -d/2,\nonumber \\
&=&-{k_x\over \vep_z k_0}(Ae^{ik_xx}-Be^{-ik_xx}),\quad -d/2\leq x\leq d/2,\nonumber \\
&=&-{i\k_0\over k_0}e^{-\k_0(x-d/2)},\quad x\geq d/2.
\label{Ez-1}
\eea

The matching of the tangential electric and magnetic fields at boundary $x=\pm d/2$ 
leads to the following eigen equation for the TM$_m$ modes
\be
k_0d={1\over \sqrt{\vep_z}\sqrt{1-n_p^2/\vep_x}}\Big(m\pi+2\arctan
{\sqrt{\vep_z}\sqrt{n_p^2-1}\over \sqrt{1-n_p^2/\vep_x}}\Big).
\label{eigen-eq}
\ee
Here $n_p\equiv\b/k_0$ is the phase index and $m=0,1,2,\cdots$, is the parity index
of the guided mode.

In order to facilitate the analysis of the eigen modes and subsequently
the calculation of the group index $n_g$, we introduce a parameter $\x$ with
\bea
k_x&=&k_0\sqrt{\vep_z(1-1/\vep_x)}/\sqrt{1-\x^2},\nonumber \\
\k_0&=&k_0\sqrt{1-\vep_x}\x/\sqrt{1-\x^2}.
\eea
For real $\vep_x$ and $\vep_z$, one has $0<\x<1$ for the guided modes. 
For complex permittivity, $\x$ takes complex values.
We further introduce two parameters 
\be
\varrho=1/\sqrt{-\vep_x\vep_z},\quad \s=-1/\vep_x.
\ee
Note that one has $\vep_x=-1/\s$ and $\vep_z=\s/\varrho^2$. 
Thus $\varrho$ and $\s$ can also be
used to characterize the IEM.

In term of $\x$, $\s$, $\varrho$, the eigen equation (\ref{eigen-eq}) becomes
\be
\sqrt{\vep_z}k_0d=F(\x,\varrho,\s)
\label{eigen-eq-2}
\ee
with the function $F$ defined as
\be
F={\sqrt{1-\x^2}\over \sqrt{1+\s}}
\Big(m\pi+2{\rm arctan} {\x\over \varrho}\Big).
\ee
The phase index is
\be
n_p={\sqrt{1+\x^2/\s}\over \sqrt{1-\x^2}}.
\ee

\subsection{Asymptotic solutions}

\begin{figure}[htbp]
\center{
\includegraphics [angle=0, width=7cm]{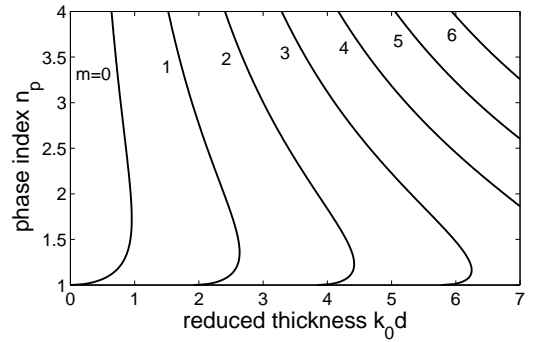}}
\caption{The phase index $n_p$
of the guided TM$_m$ modes on a free-standing planar waveguide 
of thickness $d$ with $\vep_x=-3$ and $\vep_z=2$. }
\label{fig-slab-band-np-TM}
\end{figure}

The phase index $n_p$ of the guided modes is given implicitly in the eigen equation
(\ref{eigen-eq}) or (\ref{eigen-eq-2}). 
At a fixed wavelength or frequency, $\vep_x$ and $\vep_z$ are constant. 
For real values of $\vep_x$ and $\vep_z$,
one can treat the thickness $d$ as a function of $n_p$ which is chosen as a free parameter.
The band structure for TM modes on a slab waveguide with 
$\vep_x=-3$ and $\vep_z=2$ is plotted in Fig. \ref{fig-slab-band-np-TM}.

For complex values of $\vep_x$ and $\vep_z$, the phase index $n_p$ will also be complex,
which is not so easy to solve. One must resort to numerical methods. 
Nevertheless, asymptotic solutions can be obtained analytically,
which will reveal some of the unique properties of the guided modes on IEM waveguide.

For the modes near the light line, $\x\to0$, one has the following expansion
\be
F\simeq {1\over \sqrt{1+\s}}
\Big\{m\pi+{2\x\over \varrho}\Big[1-{m\pi\varrho\over 4}\x
-\Big({1\over 2}+{1\over 3\varrho^2}\Big)\x^2\Big]\Big\}.
\ee
Thus one has
\bea
\x&\simeq& {1\over 2}\varrho(\eta k_0d-m\pi)+{m\pi\over 16}\varrho^3(\eta k_0d-m\pi)^2
\nonumber \\
&&+{1\over 16}\Big(1+{2\over 3\varrho^2}+{1\over 4}m^2\pi^2\varrho^2\Big)
\varrho^3(\eta k_0d-m\pi)^3
\eea
with $\eta=\sqrt{1+\s}\sqrt{\vep_z}=\sqrt{\vep_z}\sqrt{1-1/\vep_x}$.
The phase index for the modes near the light line is obtained as
\bea
n_p&\simeq&1+{1\over 2}(1+\s^{-1})\x^2-{1\over 4}(3-\s^{-1}+\s^{-2})\x^4
\nonumber \\
&\simeq& 1+{1\over 8}\varrho^2(1+\s^{-1})(\eta k_0d-m\pi)^2
\nonumber \\
&&+{1\over 32}(1+\s^{-1})m\pi\varrho^4(\eta k_0d-m\pi)^3\nonumber \\
&&+{1\over 64}b\varrho^4(\eta k_0d-m\pi)^4.
\eea
with the coefficient $b$ given by
\be
b=-1+3\s^{-1}-\s^{-2}
+(1+\s^{-1})({4\over 3}\varrho^{-2}+{5\over 8}m^2\pi^2\varrho^2).
\ee
So the modes near the light line is not much different from that
on a dielectric waveguide \cite{Marcuse}. Thus these modes
are forward-wave modes.

Due to the hyperbolic nature of the dispersion Eq. (\ref{hyper-disp}), 
IEM waveguide supports modes with $\k_0\to\infty$, thus $n_p\to\infty$.
There is no upper bound for $n_p$, unlike the conventional dielectric waveguide. 
For these modes, $\x\to1$, one has the expansion
\be
F\simeq {\sqrt{2(1-\x)}\over \sqrt{1+\s}}\Big(m\pi+2{\rm arctan}{1\over \varrho}\Big)
\ee
and subsequently
\be
\x\simeq 1-{(1+\s)\vep_z(k_0d)^2\over 2(m\pi+2{\rm arctan}\varrho^{-1})^2}.
\ee
It is evident that validity of $\x\to1$ requires $d\to0$. Thus one obtains
\be
n_p\simeq {\varrho\over \s k_0d}\Big[m\pi
+2{\rm arctan}{1\over \varrho}
+{\varrho^{-1}\s(1+\s)(k_0d)^2\over (1+\varrho^2)(m\pi+2{\rm arctan}\varrho^{-1})^2}\Big].
\ee
From this expression of $n_p$, it is evident that the waveguide supports infinite
number of TM modes at any thickness since all the modes with different parity
index $m$ will approach the limit $n_p\to\infty$ as $d\to0$. 
This leads to the divergence of the longitudinal electric field $E_z$ (see Eq. (\ref{Ez-1})),
similar to that on a metallic nanowire \cite{Stockman}. 
Large longitudinal electric field will be discussed more in Sec. IV.

\subsection{Energy flow and group velocity}

For waves inside the IEM, one has
\be
\b=\sqrt{\vep_x}\sqrt{k_0^2-k_x^2/\vep_z}.
\ee
If $k_x>\sqrt{\vep_z}k_0$, $\b$ will be real and negative if 
the imaginary part of the permittivity is ignored. So the waves inside the IEM
will be left-handed, $\b S_z<0$ with the $z$-component of the Poynting vector
\cite{Marcuse}
\be
S_z={1\over 2}{\rm Re}(E_xH_y^*).
\ee 
However for waves confined in the transverse direction, 
$S_z$ is no longer uniform. 
The energy flow of the guided modes is better characterized by a 
quantity $P_z$ which is defined as \cite{Marcuse}
\be
{\b\over |\b|}P_z={1\over 2}\int_{-\infty}^\infty E_xH_y^*dx. 
\ee
Here $P_z$ is actually the power per unit length (unit length in the $y$-direction).
For the TM$_m$ modes on the planar waveguide, 
$P_z$ consists of two parts,  one inside and one outside the waveguide,
\be
P_z=P_z^{\rm in}+P_z^{\rm out}\label{Pz}
\ee
with
\bea
P_z^{\rm in}&=&{|n_p|\over 4}{d\over \vep_x}\Big[1+(-1)^m{\rm sinc}\ k_xd\Big]
\nonumber \\
&=&-{|n_p|\over 4}{\s d\over F}\Big(F+{2\varrho\x\over \sqrt{1+\s}}
{\sqrt{1-\x^2}\over \varrho^2+\x^2}\Big),\nonumber \\
P_z^{\rm out}&=&{|n_p|\over 4}{1+(-1)^m\cos k_xd\over \k_0}\nonumber \\
&=&{|n_p|\over 4}{\s d\over F}{2\varrho\over \sqrt{1+\s}}
{\sqrt{1-\x^2}\over \x(\varrho^2+\x^2)}.\label{Pz-2}
\eea
The energy flow in the air $P_z^{\rm out}$ is positive as expected. 
However, due to the negative sign of $\vep_x$, the energy flow inside the waveguide is 
negative and thus contra-directional to that in the air.
A guided wave is forward (backward) wave only if $P_z$ is positive (negative).

The group velocity for the TM modes on the IEM waveguide is given by
\be
v_g={{\rm d}k_0\over {\rm d}\b}={c\over n_g}
\ee
with $c$ the speed of light in the vacuum and $n_g$ the group index defined as
\be
n_g=n_p+k_0{{\rm d}n_p\over {\rm d}k_0}. \label{ng-1}
\ee

From the above expression of $n_p$, one has
\be
{{\rm d}n_p\over {\rm d}k_0}={\pa n_p\over \pa \x}{{\rm d}\x\over {\rm d}k_0}
+{\pa n_p\over \pa \s}{{\rm d}\s\over {\rm d}k_0} \label{ng-2}
\ee
with
\bea
{\pa n_p\over \pa \x}&=&{\x\over 1-\x^2}\Big(n_p+{1\over \s n_p}\Big),\nonumber \\
{\pa n_p\over \pa \s}&=&-{\x^2\over 2n_p\s^2(1-\x^2)}. \label{ng-3}
\eea

Taking the derivative on both sides of the eigen equation (\ref{eigen-eq-2}) 
with respect to $k_0$, one has
\be
{F\over k_0}\Big(1+{k_0\over 2\vep_z}{{\rm d}\vep_z\over {\rm d}k_0}\Big)
=\pa_\x F{{\rm d}\x\over {\rm d}k_0}
+\pa_\varrho F{{\rm d}\varrho\over {\rm d}k_0}
+\pa_\s F{{\rm d}\s\over {\rm d}k_0}. \label{ng-4}
\ee
Thus one obtains
\be
{{\rm d}\x\over {\rm d}k_0}={1\over \pa_\x F}\Big[{F\over k_0}\Big(1+{k_0\over 2\vep_z}
{{\rm d}\vep_z\over {\rm d}k_0}\Big)
-\pa_\varrho F{{\rm d}\varrho\over {\rm d}k_0}
-\pa_\s F{{\rm d}\s\over {\rm d}k_0}\Big] \label{ng-5}
\ee
with
\bea
\pa_\x F&=&
{2\varrho\over \sqrt{1+\s}}{\sqrt{1-\xi^2}\over \varrho^2+\xi^2}
-{\xi F\over 1-\xi^2},\nonumber \\
\pa_\varrho F&=&-{2\over\sqrt{1+\s}}{\x\sqrt{1-\x^2}\over \varrho^2+\x^2},\nonumber \\
\pa_\s F&=&-{F\over 2(1+\s)}. \label{ng-6}
\eea

After inserting the expressions (\ref{ng-2})-(\ref{ng-6}) into Eq. (\ref{ng-1}), 
the final expression for the group index $n_g$ can be obtained.

The expression for $v_g$ is very involved. 
However, the group velocity of the modes near light line 
or with large phase index can be easily evaluated.
From the asymptotic solutions we have derived in the previous subsection, 
one can see that the guided modes near the light line are
fast modes with group velocity close to that in the vacuum. 
For the modes with very large phase index $n_p$, the group index is also large,
leading to the vanishing of the group velocity.
Due to the finiteness of the material dispersion, the divergence of $n_g$
is related to the vanishing of $\pa_\x F$ which is determined by the
geometry of the waveguide. One can also verify that
$\pa_\x F=0$ will lead to $P_z=0$ by using the expressions in Eq. (\ref{Pz-2}).

\begin{figure}[htbp]
\center{
\includegraphics [angle=0, width=8cm]{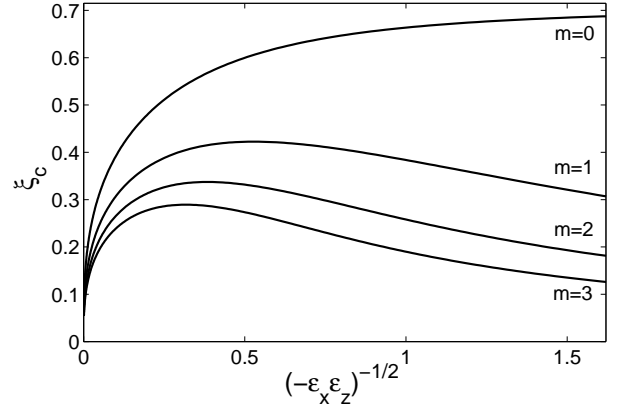}}
\caption{The dependence of the critical value $\x_c$ for the zero group velocity 
mode on the IEM parameter $\varrho=1/\sqrt{-\vep_x\vep_z}$ for the TM modes 
with different parity $m$.}
\label{fig-critical-xi}
\end{figure}

Using the above expression of $P_z$, one can determine the nature of the modes.
For the modes near the light line, $n_p\to 1$ and $\x\to0$, one has 
$P_z^{\rm in}\simeq -\s d/4$ and $P_z^{\rm out}\to+\infty$. 
Thus $P_z>0$ and the modes are forward waves.
For the modes with $n_p\to\infty$ and $\x\to1$ which requires $d\to0$, 
one has $P_z^{\rm in}\simeq -\infty$
and $P_z^{\rm out}\to\varrho^2/2k_0(\varrho^2+1)$. 
Thus $P_z<0$, the modes are backward waves. 
For modes with  finite $n_p$ and $n_p>1$, $P_z$ is finite.
From the band structure shown in Fig. \ref{fig-slab-band-np-TM},
one would expect that $P_z$ will decrease with increasing $n_p$. 
At certain critical point $d_c$ or $n_{p,c}$, $P_z=0$. Further increasing $n_p$,  
$P_z$ will become negative, indicating the mode will be backward wave.
The critical values of the phase index $n_{p,c}$ and thickness $d_c$ are determined by
a critical value $\x_c$ which satisfies the condition
\be
{2\varrho(1-\x_c^2)\over \x_c(\varrho^2+\x_c^2)}=m\pi+2{\rm arctan}{\x_c\over \varrho}.
\ee
The condition of zero group velocity is only determined 
by the parameter $\varrho=1/\sqrt{-\vep_x\vep_z}$.
The above condition can be recasted as
\be
\x_c^2=\Big[1+{1\over 2}\Big({\varrho\over \x_c}+{\x_c\over \varrho}\Big)
\Big(m\pi+2{\rm arctan}{\x_c\over \varrho}\Big)\Big]^{-1}.
\ee
The dependence of the critical values $\x_c$ on $\varrho$ 
is plotted in Fig. \ref{fig-critical-xi}.
Once $\x_c$ is determined, the critical values of $d_c$ and $n_{p,c}$
can be calculated for a given value of $\s$.

From the plotted example (Fig. \ref{fig-slab-band-np-TM}), 
a few salient features are evident. 
Unlike a conventional dielectric slab waveguide which supports 
only forward-wave modes, most modes here are backward-wave modes 
since ${\rm d}n_p/{\rm d}d<0$. Only for modes near the light line where $n_p\sim 1$, 
they are forward-wave modes since ${\rm d}n_p/{\rm d}d>0$.
As shown in Ref. \cite{Huang08}, one can prove that $\pa_\x F\geq 0$
will lead to ${\rm d}n_p/{\rm d}d\geq0$, subsequently
$P_z\geq0$ and vice versa. Thus one can use the sign of 
${\rm d}n_p/{\rm d}d$ to determine whether a mode is a forward- or backward-wave mode.
Also unlike the DNM waveguide which can support single mode 
\cite{Tsakmakidis06}, our IEM waveguide supports infinite number of TM
modes. That is, it has a very rich band structure for the guided TM modes.
This can be an advantage if utilized properly.
Our dielectric cladded IEM waveguide is also different 
from the perfect electronic conductor (PEC) cladded IEM waveguide 
considered in Ref. \cite{Podolskiy}, where
the supported modes are all backward waves. 
The coexistence of forward- and backward-wave modes will lead to
unique properties which will be revealed in the rest of the paper.

So far we have only considered IEM without loss. 
Any realization of IEMs, which will be discussed in the next section, will have loss.
The definition of IEM will be with respect to the real part of the permittivity \cite{Mackay}.
However all the derivations and expressions will still be valid in the presence of loss in IEMs.

\section{Realization and light coupling to indefinite permittivity 
material waveguide}

Nature does allow the existence of single-negative materials with
homogeneous permittivity or permeability being negative, 
such as noble metals, polaritonic materials, and ferrites. 
If the underlying crystal is anisotropic, the plasmon frequency can be anisotropic,
leading to indefinite permittivity. 
One example is bismuth in the THz range \cite{Podolskiy}, 
though bismuth must be placed in liquid helium temperatures
in order to have very small loss. 

Extremely anisotropic materials can be realized easily 
in a broad range of frequencies by proper design of metamaterials.
One method is through the engineering of artificial electric resonances. 
Another method is through the homogenization \cite{Sihvola,Mackay}.
For example for a multilayered structure of dielectric $\vep_a$ and metal $\vep_m$,
the effective permittivities can be obtained by using
the effective medium theory \cite{Sihvola,Lu07}, 
\bea
\vep_x&=&f\vep_m+(1-f)\vep_a,
\nonumber \\
\vep_z&=&{\vep_a\vep_m\over f\vep_a+(1-f)\vep_m}.\label{EMT-1D}
\eea
Here $f$ is the filling ratio of the metal.
For $f>f_{\rm min}\equiv \vep_a/(\vep_a-{\rm Re}\ \vep_m)$, one has ${\rm Re}\ \vep_x<0$.
For noble metals such as gold or silver, ${\rm Re}\ \vep_m<0$ in the 
near infrared and visible frequency range. For polar materials \cite{Kittel}, 
such as LiTaO$_3$ crystal, between the transverse and longitudinal 
optical-mode frequencies, 26.7 THz and 46.9 THz, 
the real part of permittivity is also negative. 
Thus if the periodicity of the multilayered structure is much less 
than the operating wavelength, these negative permittivity materials 
can be used to design IEMs in their respective frequency ranges. 

\begin{figure}[htbp]
\center{
\includegraphics [angle=0, width=7.5cm]{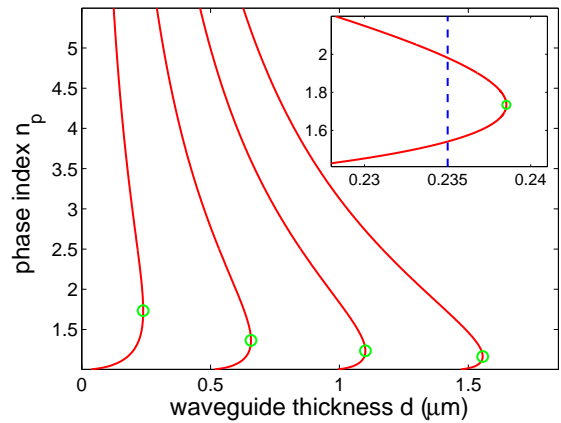}}
\caption{(Color online) The phase index $n_p$
of the first 4 TM modes on a free-standing planar waveguide 
of thickness $d$ with $\vep_x=-3.0742$ and $\vep_z=1.9762$ at $\l=1.55\m$m. 
The circle marks the location of zero group velocity.
Insert is for the TM$_0$ modes. The dashed line is for $d=235$ nm. 
The critical thickness for the TM$_0$ modes is $d_c=238.5$ nm.}
\label{fig-slab-band-np-TM-1550}
\end{figure}

We design an IEM at  $\l=1.55\m$m. 
The metamaterial is formed by using alternative layers of silver and MgF$_2$.
At this wavelength, one has $\vep_m=-129+3.3i$ \cite{Johnson,Oulton} 
and $\vep_a=1.9$ \cite{Palik}. 
Using Eq. (\ref{EMT-1D}) with  filling ratio $f=3.8\%$, we have 
$\vep_x=-3.0742+0.1254i$ and $\vep_z=1.9762+0.00003i$.
This IEM can be used to make waveguides of subwavelength thickness.
When loss is present, the permittivity components in Eq. (\ref{IEM-xz}) 
are replaced by the real parts of the permittivity.
All of our theoretical formulas in the previous section are still valid if complex 
$\vep_x$ and $\vep_z$ are used, which lead to guided modes with complex wave number $\b$
and phase index $n_p$. Solutions must be sought numerically.
If loss is ignored at this point, the guided TM modes can be easily calculated and
are plotted in Fig. \ref{fig-slab-band-np-TM-1550}.

Due to its subwavelength thickness, coupling light into an IEM waveguide 
is a technical challenge. 
Currently there are 4 ways to couple light to waveguides \cite{Reed}: 
butt coupling, end-fire coupling, prism coupling, and grating coupling. 
The coupling efficiency depends on the coupling method and the optical 
properties of the waveguide.
The end-fire coupling is a butt coupling with a focal lens. 
Multiple modes will be excited in the IEM waveguide by butt coupling. 
However for a waveguide made of realistic metamaterial, loss is unavoidable, 
thus only one or two modes will survive over certain distance and eventually only one mode 
will survive after a certain distance.

Though an IEM waveguide supports infinite number of modes, selective 
excitation of a single mode is possible. Thus one can take the full advantage of the rich 
band structure provided by the IEM waveguide.
In order to excite a single mode, the prism coupling or grating coupling should be used. 
However the phase-match condition must be satisfied for maximum energy transfer 
from the light source to the waveguide.

For a simple illustration, we use the prism coupling to excite the TM modes 
in a slab waveguide made of IEM at $\l=1.55\m$m. 
For simplicity, we first ignore the imaginary part of the permittivity.
Dissipation will be considered later in this section.
In the range $1.4<n_p<2.2$ and the slab thickness $d$ between 225 nm and 240 nm, 
only the TM$_0$ modes will be excited as shown in Fig. \ref{fig-slab-band-np-TM-1550} 
(see Fig. \ref{fig-slab-band-np-TM} for band structure of similar parameters). 
The critical thickness, such that the forward TM$_0$ and backward TM$_0$ will be 
merged into a single mode of zero group velocity, 
is $d_c=238.5$ nm with $n_p=1.734$. At the thickness $d=235$ nm, two TM$_0$ modes are allowed,
with $n_p=1.541, 1.981$. The first one is a forward-wave mode while the second is a backward-wave mode.

\begin{figure}[htbp]
\center{
\includegraphics [angle=0, width=8.5cm]{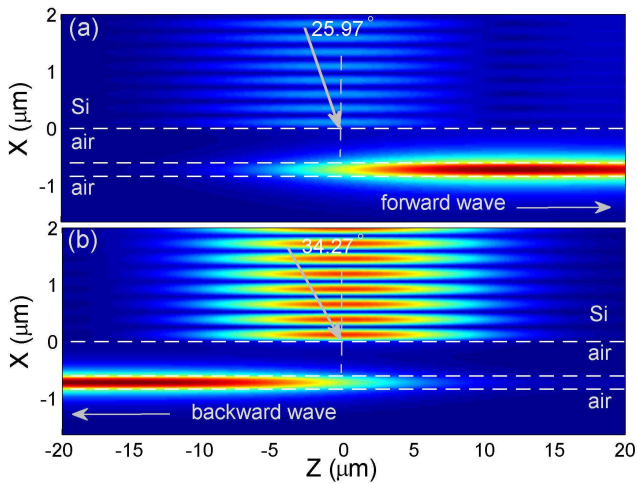}}
\caption{(Color online) Gaussian beam excitation through prism coupling 
of the forward-wave (a) and backward-wave mode (b) 
on an IEM waveguide at incident angle 25.97$^\circ$ and 34.27$^\circ$, respectively. 
The IEM has lossless permittivity $\vep_x=-3.0742$ and $\vep_z=1.9762$ at $\l=1.55\m$m.
The air gap between the prism and the waveguide ($d=235$ nm) is 600 nm.
Plotted is the absolute value of the magnetic field $H_y$.}
\label{fig-prism-coupling}
\end{figure}

We place a silicon prism of refractive index $n=3.518$ next to the IEM waveguide. The air gap
between them is 600 nm. We then shine a Gaussian beam into the silicon prism.
At an incident angle 25.97$^\circ$ inside the prism, 
the forward-wave mode will be excited while at an incident angle 34.27$^\circ$, 
the backward-wave mode will be excited, which are shown in 
Fig. \ref{fig-prism-coupling}.

\begin{figure}[htbp]
\center{
\includegraphics [angle=0, width=7.5cm]{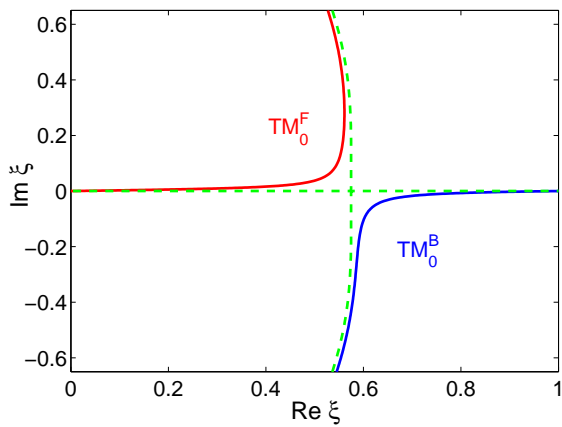}}
\caption{(Color online) The splitting of the forward-wave (red) 
and backward-wave (blue) TM$_0$ modes in the complex-$\x$ plane
in the presence of loss in the IEM with $\vep_x=-3.0742+0.1254i$ 
and $\vep_z=1.9762+0.00003i$ at $\l=1.55\mu$m. The dot-dashed line (green)
is the solution for TM$_0$ modes on the IEM waveguide without loss.}
\label{fig-TM-split-xi}
\end{figure}

In the above illustration, loss in IEM is ignored. 
When the loss is present, the eigen equation (\ref{eigen-eq-2}) will only allow
complex solutions. Every mode acquires an imaginary part. 
There is no clear cut distinction 
between propagating modes and decay modes. 
Duo to the presence of loss in the IEM, the forward-wave TM$_0^F$ modes will not be
degenerate with TM$_0^B$ at any thickness. 
This is shown in Fig. \ref{fig-TM-split-xi} in the complex-$\x$ plane.

\begin{figure}[htbp]
\center{
\includegraphics [angle=0, width=8cm]{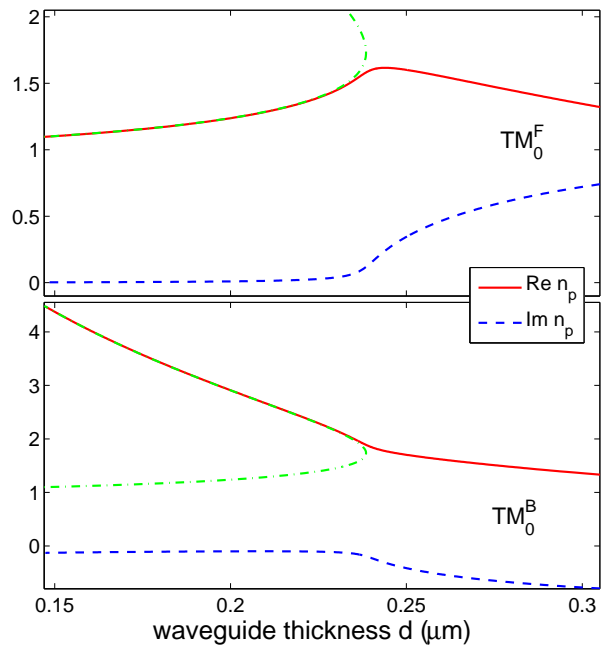}}
\caption{(Color online) The complex phase index $n_p$ for the forward-wave (top) 
and backward-wave (bottom) TM$_0$ modes
on an IEM waveguide with $\vep_x=-3.0742+0.1254i$ 
and $\vep_z=1.9762+0.00003i$ at $\l=1.55\mu$m. The dot-dashed line (green)
is the phase index of the propagation TM$_0$ modes on the IEM waveguide without loss.}
\label{fig-TM-split}
\end{figure}

The splitting of solutions in the complex-$\x$ plane will lead to the split of the phase index $n_p$ which is plotted in Fig. \ref{fig-TM-split}. 
Consequently, there is no well-defined critical thickness. 

\begin{figure}[htbp]
\center{
\includegraphics [angle=0, width=8.5cm]{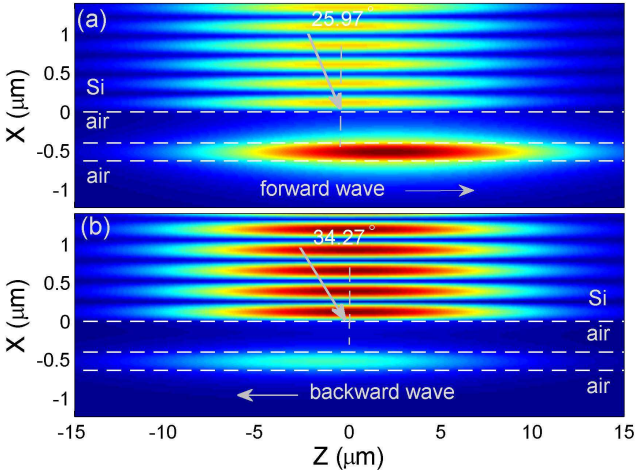}}
\caption{(Color online) Same as Fig. \ref{fig-prism-coupling} except that
the IEM has loss with $\vep_x=-3.0742+0.1254i$ and $\vep_z=1.9762+0.00003i$.
The air gap is now 400 nm.}
\label{fig-prism-coupling-loss}
\end{figure}

At a thickness $d=235$ nm, the two TM$_0$ modes have $n_p=1.5190+0.0672i$, 
$2.0026-0.1437i$ 
for the forward- and backward-wave modes, respectively.
Thus the forward- and backward-wave modes will propagate a finite distance
before lose most of their power.
Since the forward-wave mode is closer to the light line, thus the loss will have a smaller 
effect while the backward-wave modes corresponding deeper penetration into the ohmic metal,
thus will have shorter decay length.
The excitation of these two modes is illustrated in Fig. \ref{fig-prism-coupling-loss}.

\section{Applications of waveguide made of indefinite permittivity materials}

\subsection{Phase shifter and longitudinal electric field enhancement}

One salient feature of the modes on the IEM waveguide is the
large phase index. The high phase index is due to the hyperbolic dispersion 
in the metamaterial. Nanowires waveguide based on these slab waveguides
can be used for phase shifter with small footprint in optical integrated circuits.
The presence of loss in the metamaterials will restrict the use
of the waveguide. However for many applications other than the long-haul
transportation, short waveguides have the advantage of small sizes.

As examples, we consider a slab waveguide made of an IEM with 
$\vep_x=-3.0742+0.1254i$ and $\vep_z=1.9762+0.00003i$
at $\l=1.55\m$m. The realization of this IEM was discussed in the previous section.
If no loss is present, the waveguide made of this IEM will support 
forward-wave and backward-wave modes as shown in Fig. \ref{fig-slab-band-np-TM-1550}.

At a thickness $d=160$ nm, the waveguide supports two TM$_0$ modes, 
one is a forward-wave mode with $n_p=1.1211+0.0035i$ and the other is 
a backward-wave mode with $n_p=4.0298-0.1200i$.
The TM$_0^B$ with phase index ${\rm Re }\ n_p\sim 4$ can be used to create 
large phase shift over a short distance. In order to have a phase shift 
of $\pi$ by this slab waveguide, the propagating length can be about 192.3 nm. 
Over this distance, the damping is about 1.26 dB. 
For excitation of such high phase index modes, grating coupling should be used.

Besides the large phase index, modes with large longitudinal electric
field can be excited on the IEM waveguide.
Recently there is a strong interest in large longitudinal electric fields \cite{Driscoll09}.
Large longitudinal electric field can have a lot of applications,
such as superfocusing to beat the diffraction limit \cite{Dorn,Urbach}
and trapping metallic nanoparticles in optical tweezer \cite{ZhanQ}.

\begin{figure}[htbp]
\center{
\includegraphics [angle=0, width=8.5cm]{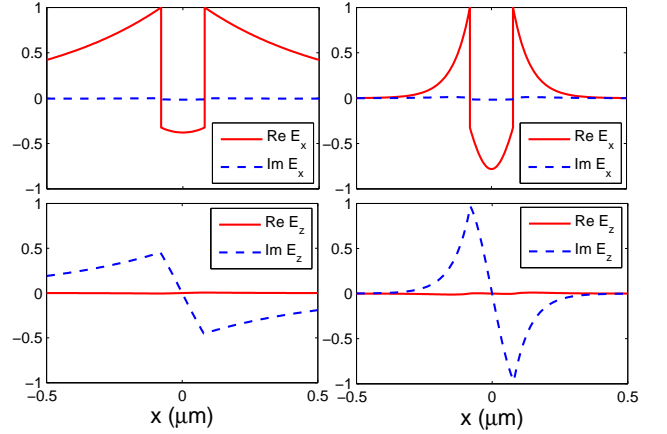}}
\caption{(Color online) The longitudinal and transverse electric fields of the 
two TM$_0$ modes on a slab waveguide
of thickness $d=160$ nm with $\vep_x=-3.0742+0.4896i$ 
and $\vep_z=1.9762+0.00003i$ at $\l=1.55\mu$m. 
One has $n_p= 1.1211+0.0035i,4.0298-0.1200i$, respectively.}
\label{fig-TM-E-field}
\end{figure}

Due to the confinement in the transverse direction, there is a $\pi/2$ phase 
difference between $E_z$ and $E_x$ in air. 
The transverse electric field is
\bea
E_x(x)&=&{\b\over k_0}e^{\k_0(x+d/2)},\quad x\leq -d/2,\nonumber \\
&=&{\b\over \vep_x k_0}(Ae^{ik_xx}+Be^{-ik_xx}),\quad -d/2\leq x\leq d/2,\nonumber \\
&=&{\b\over k_0}e^{-\k_0(x-d/2)},\quad x\geq d/2.
\label{Ex-1}
\eea
For the waveguide thickness $d=160$ nm, the electric field of
the two TM$_0$ modes are plotted in Fig. \ref{fig-TM-E-field}.

Here we consider the ratio
\be
s=|E_z|_{\max}/|E_x|_{\max}.
\ee
One can see that $s\sim 50\%$ for the TM$_0^F$ mode and $s\sim 100\%$ 
for the TM$_0^B$ mode. The ratio is comparable and even stronger than 
that on a silicon nanowire waveguide \cite{Tong04,Driscoll09}.
If one shrinks the thickness, even larger ratio would be expected. 
The fact that the IEM nanowire waveguide will have large
longitudinal electric field is due to the hyperbolic dispersion 
Eq. (\ref{hyper-disp}), which allows much stronger 
confinement of light in the transverse direction, 
thus stronger longitudinal electric field.

\begin{figure}[htbp]
\center{
\includegraphics [angle=0, width=8.2cm]{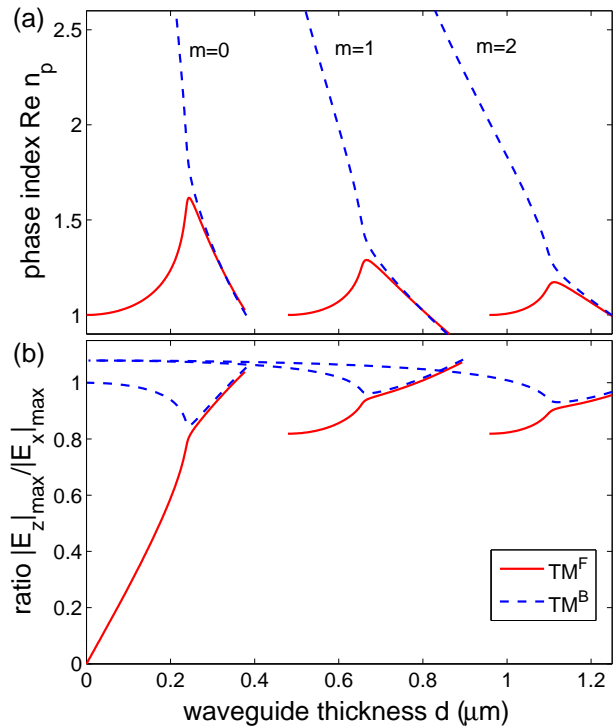}}
\caption{(Color online) (a) The real part of the complex phase index $n_p$ and (b) 
the ratio $s$ of the longitudinal and transverse electric fields of the TM modes on a free-standing
slab waveguide of thickness $d$ with $\vep_x=-3.0742+0.4896i$ 
and $\vep_z=1.9762+0.00003i$ at $\l=1.55\mu$m. }
\label{fig-E-ratio}
\end{figure}

We further derive analytical expression for the ratio $s$.
One has
\be
|E_x|_{\max}=|n_p|\max(1,|\s\sqrt{1+\x^2/\varrho^2}|).
\ee
and
\bea
|E_z|_{\max}&=&\Bigg|{\x\sqrt{1+1/\s}\over \sqrt{1-\x^2}}\Bigg|,\hskip 1.6cm  m=0,\nonumber \\
&=&\Bigg|{\sqrt{1+\s}\over \sqrt{\vep_z}}
{\sqrt{1+\x^2/\varrho^2}\over \sqrt{1-\x^2}}\Bigg|,\quad m>0.
\eea
By using the above expressions and also the eigen equation (\ref{eigen-eq-2}), 
the ratio $s$ can be calculated 
for modes with different waveguide thickness $d$
as shown in Fig. \ref{fig-E-ratio}.
Depending on the value of $\vep_x$ and $\vep_z$, the ratio $s$ can be over 100\%.

\subsection{Light wheel and open cavity formation}

Resonances are ubiquitous. They have many applications such as storing and confining 
energy, enhance the field concentration, and improve the detection accuracy. 
To have a resonance, a compact space is required, such as cavities.  
Examples are the quantum dots, microwave cavities, photonic crystal microcavities.
Recently, negative-index metamaterials are also used to form open cavities 
\cite{Notomi02,Pendry03,He05,Ruan},
such as the checkerboard open resonators \cite{Ramakrishna}.

For optical integrated circuits, one of the most difficult tasks of nanofabrication 
is the alignment of different parts and devices.
It would be more desirable to have a open cavity at any location.

Recently, a new concept, a light wheel, has been developed \cite{Tichit}. 
This is formed in a composite waveguide, which is made of an ordinary
slab waveguide coupled with a properly designed slab waveguide made of DNM.
If these two waveguides are separated infinitely away, 
the ordinary waveguide support a single forward-wave mode. 
The DNM waveguide supports a single backward-wave mode with the same phase index. 
Once these two waveguides are placed in the vicinity of each other, 
the composite waveguide no longer supports propagating modes at the same wavelength. 
Instead it will support complex-conjugate decay modes.
These two decay modes will form the so-called light wheel \cite{Tichit}.

\begin{figure}[htbp]
\center{
\includegraphics [angle=0, width=8.5cm]{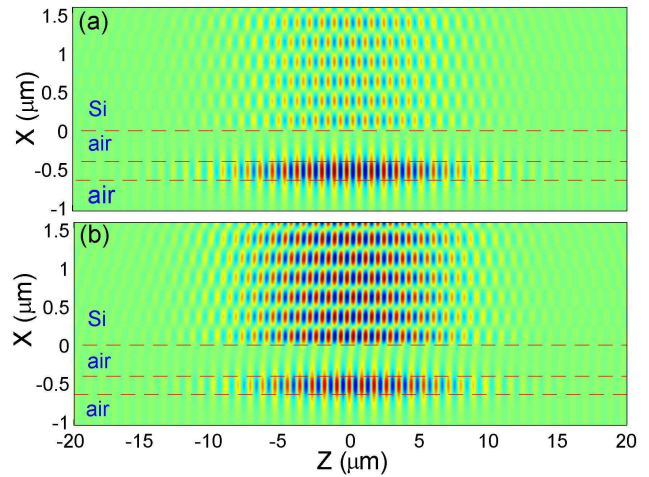}}
\caption{(Color online) Light wheel formation through a prism coupling 
on an IEM slab waveguide of thickness $d=239$ nm  at $\l=1.55 \mu$.
(a) Lossless IEM with $\vep_x=-3.0742$ and $\vep_z=1.9762$ and 
(b) IEM with $\vep_x=-3.0742+0.4896i$ and $\vep_z=1.9762+0.00003i$.
The incident angle of the Gaussian beam in the silicon prism is 29.4$^\circ$.
The air gap between the prism and the waveguide is 400 nm.
Plotted is the magnetic field $H_y$.}
\label{fig-prism-coupling-lightwheel}
\end{figure}

In order to have complex-conjugate decay modes, the waveguide should first be able 
to support degenerate propagating modes. In the example we considered in Sec. III, 
the slab waveguide is made of a metamaterial with $\vep_x=-3.0742$ and 
$\vep_z=1.9762$ at $\l=1.55 \mu$m.
Below the critical thickness $d_c=238.5$ nm, the waveguide supports two modes
of different phase indices (see Fig. \ref{fig-slab-band-np-TM-1550}). 
One mode is a forward-wave TM$_0$ mode and the other is a backward-wave TM$_0$ mode. 
At the critical thickness, the effective thickness of the waveguide is zero 
due to the negative Goos-H\"anchen lateral shift and a double light cone will be 
formed \cite{Tsakmakidis}.
However above the critical thickness, the waveguide supports no propagating modes. 
Instead, it supports complex-conjugate decay modes. 
For example at $d=239$ nm, one has
$n_p=1.7286 \pm 0.0373i$. When an incident beam with $\b=1.7286k_0$ hits the waveguide, 
the two decay modes will be excited, one decays along $\b$ and the other 
in the opposite direction of $\b$, thus an open cavity will be formed. 
This cavity can be formed at any location along the IEM waveguide, which
is shown in Fig. \ref{fig-prism-coupling-lightwheel}(a).

When dissipation is present, the decay modes no longer form a complex-conjugate pair.
However the imaginary part of their complex phase indices will still have opposite signs, 
thus light wheel formation is still allowed. In our example as shown 
in Fig. \ref{fig-TM-split},
for $d=239$ nm, the two modes has the complex phase index 
$n_p=1.5877+0.1282i,1.8730-0.2042$. The formation of light wheel is shown in 
Fig. \ref{fig-prism-coupling-lightwheel}(b).

\subsection{Slow light waveguide}

For the anisotropic waveguide we have considered, $\vep_x<0$, 
one has $P^{\rm in}_z<0$ and $P^{\rm out}_z>0$ if one sets $\b>0$ or $n_p>0$.
If $P_z=P^{\rm in}_z+P^{\rm out}_z<0$, the mode is a backward-wave mode 
since the total energy flow is 
opposite to the phase velocity. Otherwise, the mode is a forward-wave mode.
Unlike the dielectric waveguide or the PEC grating, 
the IEM waveguides support both forward and backward waves. 
At the critical thickness $d_c$, the backward and forward modes become degenerate,
the energy flow inside the waveguide cancels out that in the air. 
One can prove that at $d_c$ where $P_z=0$,
the group velocity is indeed zero. 
One does not need to
know the material dispersion to locate the zero group velocity point.
This is due to the fact that for these waveguides, 
the dispersion due to geometric confinement dominates the material dispersion at 
and around the critical thickness.

This unique property of the modes on an IEM waveguide can be used to
slow down and even trap light.
Even though the waveguide supports infinite number of TM modes
at any fixed thickness and frequency, with appropriate laser coupling,
the excitation of the TM$_m$ modes with $m\geq 1$ in the waveguide can 
be suppressed or even eliminated. 
Due to the material dissipation, 
the first TM mode will propagate the longest distance. 
The rest of the TM modes will all decay out at about half 
the decay length of the first TM mode.
It is the TM$_0$ band which can be used for slow light application, 
though other TM modes can also be used.
Unlike the double negative waveguide \cite{Tsakmakidis}, 
the IEM waveguide will slow down and trap light if one increases the thickness to 
the critical thickness $d_c$. 
These waveguides can thus be used as delay line in optical buffers \cite{XiaF07}.

For the butt-coupling, multiple modes are excited.
To excite a single mode, prism coupling or grating coupling can be used. 
By tapering the waveguide thickness, any branches of the guided modes can be accessed
in principle.

\begin{figure}[htbp]
\center{
\includegraphics [angle=0, width=7.8cm]{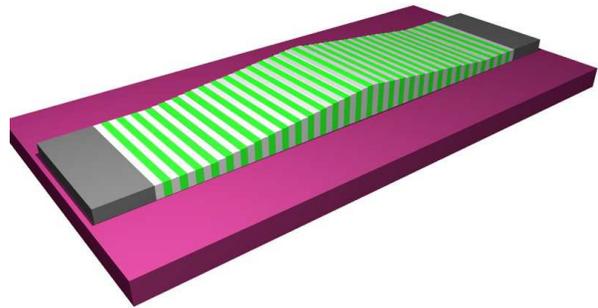}}
\caption{(Color online) A sketch of tapered slow light waveguide made 
of IEM on a glass substrate.}
\label{fig-slow-light-sketch}
\end{figure}

\begin{figure}[htbp]
\center{
\includegraphics [angle=0, width=7.5cm]{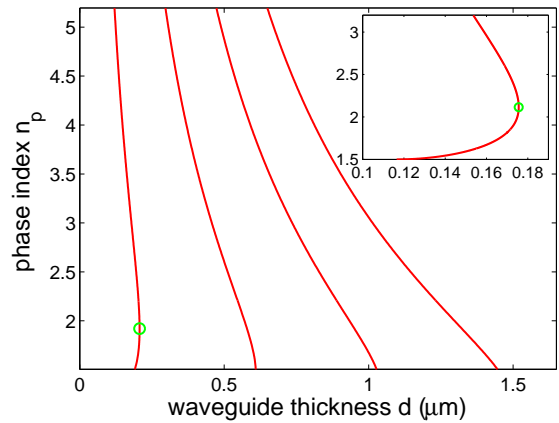}}
\caption{(Color online) The phase index $n_p$ as a function of the waveguide thickness $d$
of the first 4 guided TM modes on an IEM slab waveguide placed on a glass substrate
with refractive index $n=1.5$. 
The IEM has $\vep_x=-3.0742$ and $\vep_z=1.9762$ at $\l=1.55\ \m$m.
The circle marks the location of zero group velocity.
Insert is for the TM$_0$ modes which starts at $n_p=1.5 $ with $d=115.8$ nm. }
\label{fig-np-IEM-on-glass}
\end{figure}

For an asymmetric waveguide where an IEM waveguide is placed on a dielectric substrate,
the excitation of some unwanted TM modes can be avoided. 
A sketch of a tapered IEM waveguide on a glass substrate is illustrated in
Fig. \ref{fig-slow-light-sketch}. Let the IEM have
$\vep_x=-3.0742$, $\vep_z=1.9762$ and the glass substrate have refractive index $n=1.5$
at $\l=1.55\ \m$m.
The phase index $n_p$ as the function of thickness $d$ is plotted in 
Fig. \ref{fig-np-IEM-on-glass}.
Due to the presence of substrate, 
the phase index of the guided TM$_0$ modes starts at $n_p=1.5$ with $d_0= 116.5$ nm.
The critical thickness of the TM$_0$ modes is $d_c=174.4$ nm. Thus if the initial thickness of
a tapered IEM waveguide is less than 115.8 nm, the excitation of the forward-wave TM$_0$ mode
can be avoided. If one further ignores the excitation of TM$_m$ with $m\geq 1$ 
due to the loss or the coupling strength, the tapered waveguide
will allow only a single mode in practice. 
If the thickness is gradually increased to $d_c=175.6$ nm, the wave will be stopped there.

However, loss is ubiquitous. The presence of loss in the IEM will prevent 
forward-wave modes and the backward-wave modes from joining together 
as shown in Fig. \ref{fig-TM-split}, 
leading to finite group velocity.
For the IEM we have designed in the previous section, one has the material dispersion for the IEM
\bea
{{\rm d}\vep_z\over {\rm d}k_0}&=&{f\vep_z^2\over \vep_m^2}
{{\rm d}\vep_m\over {\rm d}k_0},\nonumber \\
{{\rm d}\varrho\over {\rm d}k_0}&=&-{f\over 2}\varrho\Big({1\over \vep_x}
+{\vep_z\over \vep_m^2}\Big){{\rm d}\vep_m\over {\rm d}k_0},\nonumber \\
{{\rm d}\s\over {\rm d}k_0}&=&{f\over \vep_x^2}{{\rm d}\vep_m\over {\rm d}k_0}.
\eea
Here we have ignored the dispersion of $\vep_a$ in Eq. (\ref{EMT-1D}).
We further assume a Drude model dispersion for silver. Thus one has
\be
{{\rm d}\vep_m\over {\rm d}k_0}={1\over k_0}\Big[2(1-\vep_m)
-i{\rm Im}\vep_m{1-\vep_m\over 1-\vep_m^*}\Big].
\ee

\begin{figure}[htbp]
\center{
\includegraphics [angle=0, width=7.8cm]{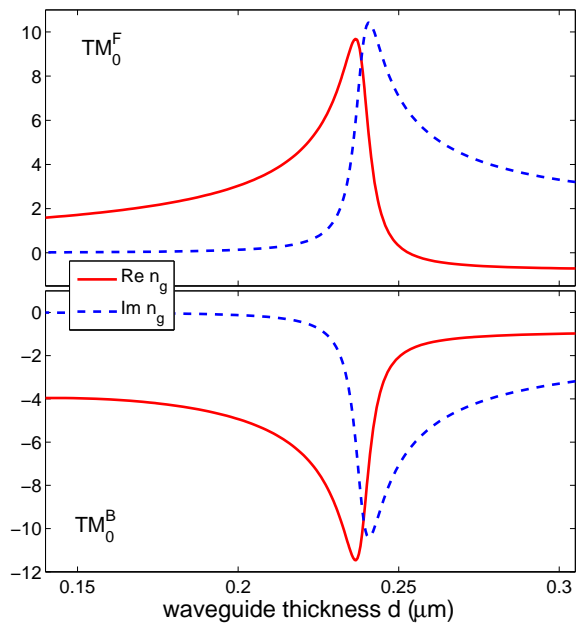}}
\caption{(Color online) Group index $n_g$ for the forward-wave (top) and 
backward-wave (bottom) TM$_0$ modes on a free-standing
IEM with $\varrho=0.4055 + 0.0083i$ and $\s=0.3247 + 0.0132i$ at $\l=1.55\ \m$m.
The dispersion of the IEM is discussed in the text.}
\label{fig-ng}
\end{figure}

For the exemplary IEM waveguide we have designed in the previous section,
the group index of the modes is calculated and plotted in Fig. \ref{fig-ng}
by using the above dispersion of the IEM. 
One can clearly see that due to the presence
of loss in the permittivity, especially in $\vep_x$, the group index is greatly reduced.
In the example we considered, we get $n_g\sim 10$. 
The fact that the group index $n_g$ for the backward-wave TM$_0^B$ modes is negative 
is due to our convention of adopting positive phase index $n_p$ throughout the paper. 

However gain can be introduced so that at a critical gain, the two branches 
of modes will recombine at a critical thickness with the same real phase index. 
One can coat gain medium such as
semiconductor optical amplifiers (SOAs) \cite{Dutta} by IEM. 
The gain in the SOA can compensate the loss in IEM
and zero group velocity mode can be recovered. Recent progress has 
also been made to use gain to boost the propagation of surface plasmon polaritons \cite{Noginov}.

We point out that the effective medium theory for the metal and dielectric multilayered
structure is still valid when one goes down in frequency from the visible, 
near infrared to THz waves and microwaves. 
The multilayered structure will still be able to
support zero group velocity modes when one decreases the frequency.
But the phase index when the group velocity is zero will increase.
In the limit of perfect electronic conductor, $\vep_m\to\infty$, one has
\be
\vep_x=\infty,\quad \vep_z=\vep_a/(1-f).
\ee
In order to give correct dispersion for waves in the metamaterial, $k_{x}=\sqrt{\vep_a}k_0$
for the TM modes (see Fig. \ref{fig-disp}), one should have 
\be
\mu_y=1-f.
\ee
A free standing slab with the above permittivity and permittivity will support guided
TM modes. Waveguide can also be formed by sandwiching this metamaterial
with dielectric on both sides, or dielectric on one side and PEC on the other.
In the case of a PEC grating with finite depth, the above effective indices will
give rise to the so-call spoof surface plasmons \cite{Pendry04}. 
Only when the phase index goes
to infinity, zero group velocity can be reached. This leads to the fact that
the supported modes are all forward-wave modes since the energy flow inside 
the PEC grating is zero and that in the air are along the phase velocity.
Though this type of waveguides can support TM modes with very small
group velocity \cite{Maier06,GanQ09}, they can not be used to stop light. 
Even in the THz range, the phase index when the group velocity is zero will be very high 
that in practice, a metallic grating structure will not be able to stop light.
Due to the hight phase index of the guided TM modes, strong reflection is expected
along a tapered metallic grating.
Only when the frequency is above the collision frequency, 
which is about 10 THz for silver \cite{Palik}, the
dispersion of a metallic grating will be hyperbolic and 
the grating structure may be used to stop light.

\section{Conclusions}

In this paper, we consider wave propagation in a slab waveguide with
anisotropic optical constant.
For extremely anisotropic cylinder where the transverse component of the permittivity 
is negative and the longitudinal is positive, 
the waveguide supports infinite number of TM modes. 
Among the supported TM modes, at most only one mode can be forward wave.
The rest of them are backward waves.

Possible realization of these IEM waveguides are proposed by
utilizing alternative layers of metal and dielectric. 
Light couplings to IEM waveguides have also been discussed.
To take full advantage of the rich band structure provided by the IEM waveguide, 
prism coupling and grating coupling can be used to selectively excite the guided modes.

Four unique properties have been revealed for the modes on slab waveguides
made of IEMs. 
The first is that the backward-wave modes can have very large phase
index. These waveguide can be used as phase shifters in optics 
and telecommunications.
The second is the large longitudinal electric field of the modes
due to the hyperbolic dispersion of the metamaterial.
The third is the formation of open cavity along in the waveguide due to its support of
complex-conjugate decay modes above the critical thickness.
The forth is that the waveguide supports modes of zero group velocity.
This is due to the fact that the waveguide can support both forward- and backward-wave
modes at a fixed thickness. 
If the waveguide is tapered, at certain critical thickness,
the two modes will be degenerate and carry zero net energy flow.
At other thickness, these waveguides support modes with small group velocity.
The presence of loss in IEMs will destroy the zero group velocity modes
but gain can be introduced to compensate loss.
These waveguides can thus be used as ultra-compact delay line in optical buffers \cite{XiaF07}.
Many more interesting features many further emerge.
The above features can lead to potential 
applications of these waveguides in optical integrated circuits.

In this paper, only indefinite $\vep$ metamaterials are considered
for waveguide applications.
Similar results can also be obtained for waveguide made of indefinite 
permeability ($\mu$) metamaterials (IMMs). 
As the case for any application of metamaterials with negative parameters,
loss will severely deteriorate the performance. 
The proposed design of IEM is by no means optimal.
Similar to the IEM waveguide, IMM waveguide will have limited application due to large loss.
A recent design of the split-ring resonators used in Ref. \cite{Shurig07} for cloaking
has very small loss of the permeability and can be used to realize 
the applications we have revealed in this work.

\section*{Acknowledgments}

We thank B. D. F. Casse and Y. J. Huang for discussions.
This work was supported by the Air Force Research Laboratories, Hanscom
through FA8718-06-C-0045 and the National Science Foundation
through PHY-0457002.

\end{document}